\documentclass[aps,pra,showpacs,amsmath,amssymb,twocolumn]{revtex4}
\usepackage{epsfig}
\usepackage{graphicx}

\begin{document}
\title{Universal scaling and pairing symmetry for high-temperature cuprate superconductors  }
\author{Huan-Qiang Zhou$^{1,2}$, Zu-Jian Ying$^{1,2,3}$, Mario Cuoco$^{2,3}$,
Canio Noce$^{2,3}$}

\affiliation{
$^{1}$ Centre for Modern Physics, Chongqing
University, Chongqing 400044, People's Republic of China\\
$^{2}$ Dipartimento di Fisica ``E. R. Caianiello'', Universit\`a
di Salerno, I-84084 Fisciano, Salerno, Italy\\
$^{3}$CNR-SPIN, I-84084 Fisciano, Salerno, Italy }

\begin{abstract}
Since the discovery of high-temperature superconductivity in the copper oxides (cuprates), intense research has been devoted to identify
universal trends among various
physical properties, aimed at providing essential information to understand the mechanisms of electron pairing and condensation. A first step in this direction was made by Uemura
and coworkers~\cite{Uemura89}, who discovered a linear relation between the zero-temperature superfluid density, $\rho _s(0)$, and the transition temperature, $T_c$. This relation works successfully in
the under-doped regime but fails in the optimally-doped and over-doped materials.
A similar trend has been uncovered for a scaling relation between $\rho _s(0)$ and the d.c. conductivity, $\sigma _{\rm DC}$, measured at  $T_c$~\cite{Pimenov99}. Recently, a remarkable relation
$\rho _s(0)\propto \sigma _{\rm DC}\ T_c$ was proposed by Homes and coworkers~\cite{Homes04}; it is claimed to work for all the cuprate superconductors, thus supporting the $d$-wave pairing symmetry. However, it has been subsequently observed that the Homes relation
breaks down at least in the over-doped regime~\cite{Tallon06}, if the data for $ \sigma _{\rm DC}$ are exploited
from the electrical resistivity measurements. This strongly suggests that  the pairing symmetry in the cuprates is not pure $d$-wave. Here,
we report a scaling relation between $\rho _s(0)$ and the product of the pseudogap temperature $T^{*}$, $\sigma
_{\rm DC}$ and $T_c$, with $T^{*}$ rescaled by the maximum transition temperature $T_c^{\max }$.  This scaling relation holds universally in the entire doping range for the hole-doped
cuprates. One of its ramifications is that the ratio $T^{*}/ T_c^{\max }$ reflects the extent to which the pairing symmetry
deviates from pure $d$-wave pairing. It also supports the idea that the pseudogap coexists with the superconducting gap in the superconducting phase.
\end{abstract}
\date{\today}
 \maketitle

%
\begin{figure}[h]
\begin{center}
\includegraphics[width=0.45\textwidth]{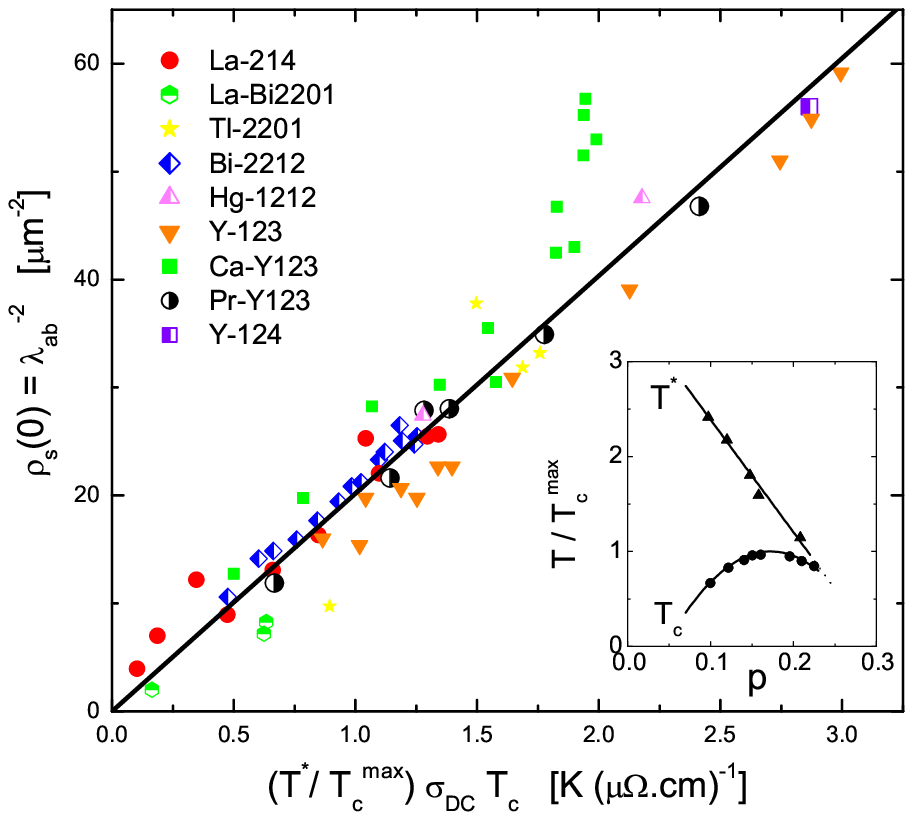}
\end{center}
\caption{ The superfluid density $\rho _s(0)$, at absolute zero, versus the
 product of the pseudogap temperature $T^{*}$, the d.c. conductivity $\sigma
_{\rm DC}$ measured at $T_c$, and the superconducting transition temperature $T_c$,  rescaled by  the maximum
transition temperature $T_c^{\max }$,  for
a variety of the hole-doped cuprates.  The data,  with a wide
hole doping range, distribute around a universal straight
line. The d.c. conductivity is determined from the $a$-$b$ CuO$_2$ plane electrical resistivity just above the
normal to superconducting transition temperature $T_c$. The London penetration depth
$\lambda_{ab}$, $\sigma _{\rm DC}$ and $T_c$ for La-214, Bi-2212,
YCa-123 and Tl-2201 are taken from Refs.~\cite{Tallon06,Tallon03}, while the data for other compounds are taken from:
Refs.~\cite{Zimmermann95Y123,Takenaka94Y123R,LeeY123R1} for Y-123,
Refs.\cite{Shengelaya98Y124,Bucher93Y124R} for Y-124,
Refs.~\cite{SeamanYPr123,Levin00} for Pr-Y123,
Refs.~\cite{RussoLaBi2201,Ando04La214andBi2201} for La-Bi2201, and
Refs.~\cite{ThompsonHg1212,KwonHg1212,ShenHg1212R} for Hg1212. Here, the
$\lambda_{ab}$ data are measured by the ac susceptibility for La-214,
the field-dependent specific heat for Bi-2212, the equilibrium
magnetization via a superconducting quantum interference device
(SQUID) for Hg-1212 and the muon spin rotation ($\mu SR$) for all the
other compounds. We take the $a$ axis $\sigma _{DC}$ values versus
the oxygen excess per CuO$_2$ plane for Y-123 and Y-124 to minimize
the chain influence along the $b$ axis. The inset illustrates the
evolution of $T^{*}$ and $T_c$ with respect to the hole doping $p$, with
Bi-2212 as a typical example. The data are taken from the
resistivity measurements~\cite{Oda97}, with $T^{*}$ being the same as
that from the nuclear magnetic resonance (NMR) measurements~\cite{Ishida98}.
Actually, the data for
$T^{*}/T_c^{\max }$, as shown in the inset, are also used for other compounds, since they
collapse between Bi-2212 and La-214~\cite{Nakano98}, Bi-2212 and
Tl-2201~\cite{Hufner08}, La-214 and
La-Bi2201~\cite{Ando04La214andBi2201}, as measured by means of different probes.
The $T^{*}/T_c^{\max }$ data for Hg-1212 are similar,
as shown in the NMR measurements~\cite{Itoh98}, if the doping $p$ is rescaled by the optimal doping value.
The evolution  of $T^{*}$ with respect to the oxygen excess in the CuO$_2$ plane for Y-123~\cite{Yasuoka97} and for
Y-124~\cite{MachiY124} are taken from the NMR measurements.
In addition, $T^{*}$ versus Pr doping for Pr-Y123 is from
the resistivity measurements~\cite{Levin00}. } \label{fig1}
\end{figure}
%

The scaling relation, which holds universally for a variety of the
hole-doped cuprates, takes the form:
\begin{equation}
\rho _s(0)\equiv\lambda _{ab}^{-2}=C\frac{T^{*}}{T_c^{\max }}\
\sigma _{\rm DC}\ T_c, \label{scaling}
\end{equation}
where $\rho _s(0)$ is the superfluid density at absolute zero temperature
and $\lambda _{ab}$ represents the in-plane London
penetration depth, with $C \sim 0.2 \; \Omega \;(K \mu m)^{-1}$ being a
universal constant. The transition from a normal state to a
superconducting state occurs at the transition temperature $T_c$, usually with a narrow
transition width. $T_c^{\max }$ is the maximum transition temperature at the optimal doping level. The d.c. conductivity $\sigma
_{\rm DC}$ is extracted from the in-plane resistivity  $ \rho _{ab}$ measured just above the
normal to superconducting transition temperature $T_c$. The pseudogap
temperature $T^*$ is defined to be the temperature, below which
the resistivity drops below its linear temperature dependence at high
temperature~\cite{Oda97}. The nuclear magnetic
resonance (NMR) relaxation rate divided by temperature $T$, $1/T_1 T$,
also takes a maximum value at $T^*$~\cite{Ishida98}.

The scaling relation is plotted in Fig.\;\ref{fig1} for
different families of the hole-doped cuprate superconductors:
Bi$_2$Sr$_2$CaCu$_2$O$_{8+\delta}$ (Bi-2212),
La$_{2-x}$Sr$_x$CuO$_4$ (La-214), Tl$_2$Ba$_2$CuO$_{6+\delta}$
(Tl-2201), YBa$_2$Cu$_3$O$_{7-\delta}$ (Y-123),
Y$_{1-x}$Ca$_x$Ba$_2$Cu$_3$O$_{7-\delta}$ (Ca-Y123),
YBa$_2$Cu$_4$O$_8$ (Y-124), Bi$_2$Sr$_{2-x}$La$_x$CuO$_{6+\delta}$
(La-Bi2201), Y$_{1-x}$Pr$_x$Ba$_2$Cu$_3$O$_{7-\delta}$ (Pr-Y123),
and HgBa$_2$CaCu$_2$O$_{6+\delta}$ (Hg-1212). Inside the
brackets are the corresponding abbreviations for these families. The families
cover a wide hole doping range under the superconducting dome: $p\;\sim[0.05,
0.27]$, with the under-doped (over-doped) regime below (above) the
optimal doping around $p\sim 0.16$. Specifically, $p\sim[0.1,0.26]$ for Bi-2212,
$p\sim[0.06,0.24]$ for La-214, $p\sim[0.075,0.25]$ for YCa-123,
$p\sim[0.13,0.26]$ for Tl-2201, $p\sim[0.13,0.185]$ for La-Bi2201,
and up to optimal doping for other families. Moreover, the
number of the copper oxide layers per unit cell varies from one (La-214, La-Bi2201,
Tl-2201) to two (Bi-2212, Hg-1212, Y-123, Ca-Y123, Pr-Y123,Y-124).
As one sees in Fig.\;\ref{fig1}, all the data, for compounds with a wide doping range and varied layer structures, collapse onto one single
straight line. That is, the universal scaling relation (\ref{scaling}) is valid.
We remark that the d.c. conductivity is determined from
the CuO$_2$ plane electrical resistivity just above $T_c$. The London penetration depth
$\lambda_{ab}$, $\sigma _{\rm DC}$ and $T_c$ for La-214, Bi-2212,
YCa-123 and Tl-2201 are taken from Refs.~\cite{Tallon06,Tallon03}, while the data for other compounds are taken from
Refs.~\cite{Zimmermann95Y123,Takenaka94Y123R,LeeY123R1} for Y-123,
Refs.\cite{Shengelaya98Y124,Bucher93Y124R} for Y-124,
Refs.~\cite{SeamanYPr123,Levin00} for Pr-Y123,
Refs.~\cite{RussoLaBi2201,Ando04La214andBi2201} for La-Bi2201, and
Refs.~\cite{ThompsonHg1212,KwonHg1212,ShenHg1212R} for Hg1212. Here, the
$\lambda_{ab}$ data are measured by the ac susceptibility for La-214,
the field-dependent specific heat for Bi-2212, the equilibrium
magnetization via a superconducting quantum interference device
(SQUID) for Hg-1212 and the muon spin rotation ($\mu SR$) for all the
other compounds. We take the $a$ axis $\sigma _{DC}$ values versus
the oxygen excess per CuO$_2$ plane for Y-123 and Y-124 to minimize
the chain influence along the $b$ axis. The inset illustrates the
evolution of $T^{*}$ and $T_c$ with respect to the hole doping $p$, with
Bi-2212 as a typical example. The data are taken from the
resistivity measurements~\cite{Oda97}, with $T^{*}$ being the same as
measured in the NMR experiments~\cite{Ishida98}.
Actually, the data for
$T^{*}/T_c^{\max }$, as shown in the inset, are also used for other compounds, since they
collapse between Bi-2212 and La-214~\cite{Nakano98}, Bi-2212 and
Tl-2201~\cite{Hufner08}, La-214 and
La-Bi2201~\cite{Ando04La214andBi2201}, as measured by means of different probes.
The $T^{*}/T_c^{\max }$ data for Hg-1212 are similar,
as shown in the NMR experiments~\cite{Itoh98}, if the doping $p$ is rescaled by the optimal doping value.
The evolution  of $T^{*}$ with respect to the oxygen excess in the CuO$_2$ plane for Y-123~\cite{Yasuoka97} and for
Y-124~\cite{MachiY124} are taken from the NMR measurements.
In addition, $T^{*}$ versus Pr doping for Pr-Y123 is from
the resistivity measurements~\cite{Levin00}.

We stress that, the
deviation, as observed in Fig.\;\ref{fig1} for some samples for YCa-123 (Ca doping $x=0.2$),  may be attributable to the fact that the d.c. conductivity data are extracted from thin film samples~\cite{Tallon06}. Indeed, it is expected that the
d.c. conductivity, if available, would be larger for single crystal samples. The
YCa-123 data will be driven closer to our scaling line, as one may judge from an
estimation in terms of the available d.c. conductivity data for single crystal samples with
Ca doping $x=0.12,0.14$~\cite{NagasaoYCa123crystal} by assuming that the d.c.
conductivity for Ca doping $x=0.2$ exhibits
a similar behavior.

The justification of our scaling relation, unveiled in this Letter,  resides in the pairing symmetry: the
Homes scaling relation, if
it were valid for high-temperature cuprate superconductors, would
be convincing evidence for pure $d$-wave pairing symmetry for the
cuprates~\cite{Anderson}. However, as observed subsequently by Tallon and coworkers~\cite{Tallon06}, there is a significant deviation from the Homes relation, with a salient feature that the deviation is increasing with increasing doping. This strongly suggests that the pairing symmetry realized in the (hole-doped) cuprates is not pure $d$-wave. In fact, growing experimental
evidence~\cite{Muller96YBCOsd,MasuiRamanSplusD,BakrRamanYCa123,KhasanovLSCOsd,Khasanov07UniversalSD,Furrer08sd,Strohm,Nemetschek,Lu,Uchiyama,Kirtley}
points to an admixture of an $s$-wave component to the predominant $d$-wave superconductivity in the hole-doped cuprate superconductors.  In addition,
an unbiased numerical simulation of the two-dimensional $t-J$ model has
demonstrated that the pairing symmetry is of $d+s$-wave nature,  with the $s$-wave component
increasing with increasing doping~\cite{Zhou,Zhou1}, consistent with the
electronic Raman scattering experiments~\cite{Nemetschek,MasuiRamanSplusD}. Here, we emphasize that such an admixture arises from the fact that the tetragonal lattice symmetry is broken locally, as evidenced by the simulation of the $t-J$ model~\cite{Zhou,Zhou1}. This in turn explains why a $d + s$-wave
superconductivity model works so well in the hole-doped cuprates~\cite{Nemetschek}.   Therefore, it is necessary to remove an extra contribution
from the $s$-wave superconducting component to the superfluid density. Equivalently, a proper weight that accounts for the contribution to the $d$-wave component should be included in the normal state d.c. conductivity $\sigma _{\rm DC}$.  Remarkably, this weight tracks the ratio of the pseudogap temperature $T^*$ and the maximum transition temperature $T_c^{\max }$. That is,
the ratio, $T^*/T_c^{\max }$, reflects the extent to which the pairing symmetry
deviates from pure $d$-wave pairing. It is worth mentioning that the pseudogap temperature $T^*$  plays a similar role in a two-fluid description of the under-doped cuprate superconductors~\cite{Pines}: $T^*$ tracks the weight of the spin liquid component versus a non-Landau Fermi liquid.

Last but not least, our results lend further support to the picture that the pseudogap coexists with the superconducting gap in the superconducting phase~\cite{Hufner08}, due to the fact that the scaling relation simultaneously involves both $T^*$ and $T_c$, two important temperature scales that dominate the phase diagram of the high-temperature
cuprate superconductors.

Helpful discussions with Carmine Attanasio are acknowledged. Huan-Qiang Zhou would like to thank the Dipartimento di Fisica, Universit\`a di Salerno for hospitality during his stay.


\begin{references}
\bibitem{Zhou1} Zhou, H.-Q. Pairing mechanism for high temperature superconductivity in the cuprates: what can we learn from the two-dimensional $t-J$ model? arXiv:1001.3358.
\end{references}
\end{document}